\documentclass[12pt,a4paper]{article}
\usepackage[a4paper]{geometry}

\usepackage{amsmath}
\usepackage{amsfonts}
\usepackage{amssymb}

\usepackage{mathrsfs}
\usepackage[T1]{fontenc}
\usepackage{mathpazo}
\usepackage{latexsym}
\usepackage{color}
\usepackage{nicefrac}
\usepackage[latin1]{inputenc}
\usepackage[nosort]{cite}
\usepackage{multirow}
\usepackage{url}

\def\be{\begin{equation}}
\def\ee{\end{equation}}
\def\ba{\begin{eqnarray}}
\def\ea{\end{eqnarray}}

\newcommand{\op}{\hspace{1pt}}
\newcommand{\eq}[1]{\begin{equation}
                     \begin{split} #1 \end{split}
                     \end{equation}}
\newcommand{\cc}{\mathsf }

\numberwithin{equation}{section}

\begin{document}

\thispagestyle{empty}
\renewcommand{\baselinestretch}{1.2}

\vspace*{-1,5cm}
\begin{flushright}
  {\small
    LMU-ASC 12/19\\
    MPP-2019-55
  }
\end{flushright}

\vspace{0.5cm}

\begin{center}
{\LARGE
    \textbf{Open-String Non-Associativity in \\[5pt]  an  R-flux Background}
}
\end{center}

\vspace{0.2cm}

\begin{center}
\renewcommand{\thefootnote}{\fnsymbol{footnote}}

\begin{small}
     \textbf{Dieter L\"ust}\footnote{dieter.luest@lmu.de}${\op}^{1,2}$,\:
     \textbf{Emanuel Malek}\footnote{emanuel.malek@aei.mpg.de}${\op}^3$,\:
     \textbf{Erik Plauschinn}\footnote{erik.plauschinn@lmu.de}${\op}^2$,\:
     \textbf{Marc Syv{\"a}ri}\footnote{marc.syvaeri@physik.uni-muenchen.de}${\op}^{1,2}$
\end{small}

\renewcommand{\thefootnote}{\alph{footnote}}
\setcounter{footnote}{0}
\end{center}

\vspace{0.2cm}

\hspace{1.25cm}\begin{minipage}{.8\linewidth}
{\it \begin{footnotesize}
\begin{itemize}

\item[${}^1$] Max-Planck-Institut f\"ur Physik,\\
F\"ohringer Ring 6, 80805 M\"unchen, Germany

\item[${}^2$] Arnold-Sommerfeld-Center f\"ur Theoretische Physik,\\
Department f\"ur Physik, Ludwig-Maximilians-Universit\"at M\"unchen,\\
Theresienstra\ss e 37, 80333 M\"unchen, Germany

\item[${}^3$] Max-Planck-Institut f\"{u}r Gravitationsphysik (Albert-Einstein-Institut), \\
Am M\"{u}hlenberg 1, 14476 Potsdam, Germany

\end{itemize}
\end{footnotesize}}
\end{minipage}

\vspace{0.5cm}


\begin{center}
\textsc{Abstract} 
\end{center}
We derive the commutation relations for  open-string coordinates on  D-branes in non-geometric 
background spaces. Starting from D0-branes on a three-di\-men\-sional torus with $H$-flux,  
we show that open strings with end points on D3-branes in a three-dimensional $R$-flux 
background exhibit a non-associative phase-space algebra, which is similar to the non-associative 
$R$-flux algebra of closed strings.
Therefore, the effective open-string gauge theory on the D3-branes is expected to be a non-associative 
gauge theory. We also point out differences between the non-associative 
phase space structure of open and closed strings in non-geometric backgrounds, which are 
related to the different structure of the world-sheet commutators of open and closed strings.


\clearpage
\tableofcontents


\section{Introduction}

It has been known for some time, 
that the position coordinates of strings 
do not commute among each other in the presence of certain Neveu-Schwarz--Neveu-Schwarz (NS-NS) background fields.
It follows that string coordinates in general cannot be simultaneously measured if appropriate 
generalized magnetic background fields are turned on.
Non-commutative string geometry was first discovered for 
matrix models and for
open strings \cite{Connes:1997cr,Douglas:1997fm,Chu:1998qz,Schomerus:1999ug,Seiberg:1999vs} (for a comprehensive review on non-commu\-tative field theories see \cite{Szabo:2001kg})
with their  endpoints on D-branes
with a non-vanishing Kalb-Ramond $B$-field or gauge-field background turned on along the D-brane world-volume.
A prototypical example of open-string non-commu\-tativity are open strings with endpoints on D2-branes plus a non-vani\-shing $B$-field
along the D2-brane directions. From the world-sheet point of view, the $B$-field induces a mixing between 
Neumann (N) and Dirichlet (D) boundary
conditions for the open-string coordinates along the D2-brane directions, and as a result  an open-string world-sheet CFT computation
yields the following commutator for the endpoints of the open-string coordinates 
\begin{equation}\label{openbasic}
\begin{split} 
	\left[ \cc X^i, \cc X^j\right]_{\rm open}=2\pi \alpha' i \op\theta^{ij}
	\qquad{\rm with}\qquad \theta^{ij}=
	- \left(\frac{1}{g+ B} \op B\op\frac{1}{g- B}\right)^{ij}\,.
\end{split}
\end{equation}
Here $\theta^{ij}$ is the open-string non-commutativity parameter.
This non-commu\-ta\-tive open-string algebra corresponds to a {\sl Poisson structure}.
As it is well-known, performing one T-duality along the  D2-brane world volume direction, one obtains an open string on a D1-brane, which lies at a certain angle in the compact space.
Now the open-string boundary conditions are Neumann along the D1-brane and Dirichlet in the perpendicular direction, and therefore the 
coordinates are commutative, as the original $B$-field on the D2-brane is T-dualized into the angle parameter of the dual D1-brane.
A second remark concerns non-constant $B$-field configurations. In this case, in general, the Poisson structure no longer exists, and it is possible that the algebra has a non-zero Jacobiator and thus becomes non-associative, as was shown in \cite{Cornalba:2001sm, Herbst:2001ai}.

About ten years after the first open-string analysis, a similar situation was found for closed strings:
closed strings in certain non-geometric NS-NS backgrounds (for a recent review  on non-geometric backgrounds in string theory see \cite{Plauschinn:2018wbo}) behave in a non-commutative or even non-associative way 
\cite{Blumenhagen:2010hj,Lust:2010iy,Blumenhagen:2011ph,Blumenhagen:2011yv,Lust:2012fp,Bakas:2013jwa}.
Namely in a so-called $Q$-flux background, the closed-string coordinates also possess some kind of mixed
closed-string boundary conditions and do not commute among each other. Performing a world-sheet calculation one can show that the commutator between
two closed-string coordinates is
 determined by a closed-string winding number 
  \cite{Lust:2010iy,Condeescu:2012sp,Andriot:2012vb,Blair:2014kla,Bakas:2015gia}:
\begin{equation} \label{rel_001}
\begin{split} 
	\bigl[X^i,X^j\bigr]_{\rm closed}=
	i\op  Q_k{}^{ij}\op\tilde p^k\,.
\end{split}
\end{equation}
Due to the appearance of the winding number $\tilde p^k$ on the r.h.s. of this equation, 
closed-string non-commutativity is a non-local effect.
The non-locality arises since the $Q$-flux background can be viewed
as a metric and $B$-field background that is globally not well-defined but rather needs a T-duality transformation 
to be patched together in a globally consistent way, which is only consistent in string theory but not in field theory.

Going one step further, 
{so-called} non-geometric $R$-flux backgrounds lead to
a non-associative closed-string algebra structure.\footnote{These non-commutative and non-associatives structures also appeared in more mathematics oriented
literature  \cite{Bouwknegt:2000qt,Bouwknegt:2004ap,Mathai:2004qq,Mathai:2004qc}, where fibrations  are applied to characterize
these kind of non-geometric backgrounds with D-branes and $B$-fields.}
Here the non-geo\-metric flux $R$ is 
the parameter that
controls 
the violation of the Jacobi identity, i.e. $R$ is the deformation parameter of the non-associative algebra of the 
$R$-flux backgrounds.
Concretely the closed-string non-commutativity in the $R$-flux background 
takes the form
\begin{equation}\label{rcom}
\begin{split} 
	\bigl[X^i,X^j\bigr]_{\rm closed}=
	i\op R^{ijk}\op p_k\,,
\end{split}
\end{equation}
where
the momentum operator $p_k$ appears on the r.h.s. of the equation. This behaviour can be understood from noting that the $R$-flux background is even locally not a well-defined manifold but needs a T-duality transformation at every "point" to be locally patched together. From the world-sheet point of view, the $R$-flux space can be regarded as a completely left-right-asymmetric non-geometric string background.
The non-associativity  of the $R$-flux background eventually arises {after} taking into account the non-vanishing canonical commutator between the position and momentum operator,
which together with \eqref{rcom} leads to a non-vanishing three-bracket of the following form:
\eq{
\label{asso} 
&\bigl[X^k,p_k\bigr]=i \,,
\\[4pt]
&\bigl[ X^i, X^j, X^j\bigr]_{\rm closed}:=\bigl[ [X^i, X^j], X^j\bigr]_{\rm closed}\pm\,{\rm perm.}\simeq  
R^{ijk}\, .     
}
These two equations now define a so-called {\sl twisted Poisson structure}. 
As discussed in \cite{Mylonas:2012pg}, this algebra can also be nicely derived  by quantizing an associated membrane sigma model.
Note that {to obtain} 
the above algebra, it is crucial that $X^i$ and $p_i$ are canonically-conjugate variables.
As we will furthermore explain, 
closed-string coordinates $X^i$ and their dual coordinates $\tilde X_i$ do not obey a simple canonical commutation relation of this form, namely we will see that
(see also with the work of \cite{Freidel:2017wst,Freidel:2017nhg}): 
\begin{equation}
\bigl[X^i,\tilde X_j\bigr]_{\rm closed}=0\neq i\op \delta^i{}_j\, .
\end{equation}

In this note we derive in a concrete setting a non-associative algebra quite  similar to \eqref{rcom} and \eqref{asso}, but now for open-string
coordinates in non-geometric 
 backgrounds.
To derive the relevant algebra we consider open string  world-sheet commutators, as they are already known from the previous work with constant $B$-field,
but now also applied to
the case of a non-constant $B$-field background. We will discuss the open-string non-commutativity and non-associati\-vity in
the context of  (non)-geometric $H$-, $F$-, $Q$- and $R$-flux backgrounds, which are related by a chain of T-duality transformations.
T-duality transformations for open strings with Neumann and Dirichlet boundary conditions on non-geometric backgrounds were discussed before in \cite{Blumenhagen:2000wh,Blumenhagen:2000fp}, and the  purpose of this work is to add the derivation of the world-sheet commutation relations of the open-string coordinates and to explore the effects on non-associativity.

It was already observed \cite{Cornalba:2001sm, Herbst:2001ai} that for non-constant $B$-field configurations the non-commutative open-string algebra no longer admits a Poisson structure in general. Instead, the  non-commutative algebra has a non-zero Jacobiator and is non-associative. 
The non-associative open-string structure that we will investigate in this work will not appear in the presence of the $H$-field background, but instead for D3-branes in the T-dual $R$-flux background.
We therefore expect that the gauge theory on the D3-branes will be a non-associative gauge theory, which possesses an $L_\infty$ structure along the lines discussed in  \cite{Blumenhagen:2018kwq}.

Concretely, our starting point are D0-branes on a background with linear $B$-field and therefore constant $H$-field. Here all three open-string coordinates are fully commutative,
meaning the open-string coordinates, i.e. the D0-brane positions, can be unambiguously measured on a torus with $H$-flux. 
After two T-dualities one obtains D2-branes on a non-commutative $Q$-flux background, where the non-commutativity parameter is just linear in the third coordinate, after using the appropriate open-string variables
provided by the Seiberg-Witten map.
This is precisely the situation which was already discussed in the work of \cite{Grange:2006es,Brodzki:2007hg} or  earlier also in \cite{Kapustin:1999di}.
After a third  T-duality 
{we obtain} an $R$-flux background, which is now fully
{wrapped}
by D3-branes. 
For this background, we uncover open string non-associativity  by deriving a non-vanishing three-bracket for the open-string coordinates in the non-geometric $R$-flux background.
Now the open string coordinates cannot be simultaneously measured anymore. In other words, there is a minimal volume on the D3-brane \cite{Mylonas:2013jha}, which also forbids the existence of probe point particles. This is consistent with the Freed-Witten anomaly \cite{Freed:1999vc}, which states that D3-branes on backgrounds with $H$-flux and hence, by T-duality, D0-branes on the non-geometric $R$-flux backgrounds are not possible. Let us mention that one can have lower-dimensional branes filling part of the three-dimensional space with $H$-flux, still being allowed by the Freed-Witten anomaly, like D2-branes on $S^3$. These give rise to a non-commutative but still associative behavior of the D2-brane gauge theory with non-constant $B$-field. Another example considered in \cite{Blumenhagen:2018kwq} are D6-branes on an $SU(3)$ group manifold with $H$-flux. Here one obtains a gauge theory on a non-commutative and also non-associative space without Poisson structure.

As we will show, open-string coordinates and their duals, unlike their closed strings counterparts, obey canonical-like commutation relations:
\begin{equation}\bigl[\cc X,\tilde {\cc X}\bigr]_{\rm open}\simeq  i\, .
\end{equation}
With this crucial ingredient, we will show that closed- and open-string non-associativity on non-geometric background geometries are governed by similar algebraic structures. However, there will also be subtle and important differences between open and closed strings, which we will mention in this  paper.

\bigskip
We finally mention that D-branes on non-geometric backgrounds together with their corresponding
non-commutative  gauge theories have 
also been studied in \cite{Hull:2019iuy}. We briefly comment on the relation between \cite{Hull:2019iuy}
and our work in the conclusions.


\section{Open string non-commutativity}


\subsection{Open strings in B-field backgrounds}

In this section, we  give a short repetition of the so-called Seiberg-Witten {map}, which enables us to go from a description of {an} open string with a $B$-field to a description without a $B$-field but with 
{a} non-commutative spacetime. The case of a constant $B$-field was first studied in \cite{Seiberg:1999vs} by Seiberg and Witten, and later generalized to cases of non-constant $B$-field in \cite{Cornalba:2001sm, Herbst:2001ai}. We are going to repeat some of their arguments in this section and start with an open string with a constant $B$-field.
The worldsheet action is 
\begin{equation}
\begin{split} 
	S=\frac{1}{4\pi \alpha^{\prime}}\int_{\Sigma} d^2\sigma \,g_{ij}\op \partial_a \cc X^i \partial^a \cc X^j
	-\frac{i}{4\pi\alpha'} \int_{\partial \Sigma} d\sigma \,B_{ij}\cc X^i\partial_t \cc X^j\,,
\end{split}
\end{equation}
where $\Sigma$ is the string worldsheet with Euclidean signature, $\partial\Sigma$ denotes the 
corresponding boundary 
and $d^2\sigma$ and $d\sigma$ denote the top-form on $\Sigma$ and $\partial\Sigma$, respectively. 
The boundary condition in the $\cc X^i$-direction on the $p$-branes then become
\begin{equation}
\begin{split} 
	\left. g_{ij}\partial_n \cc X^j + B_{ij} \partial_t \cc X^j\right|_{\partial \Sigma}=0\,,
\end{split}
\end{equation}
where $\partial_n$ is the normal derivative to $\partial \Sigma$. This condition can be conformally mapped {to the upper half-plane with coordinates $z$ and $\bar{z}$.} 
Now the boundary conditions become
\begin{equation}
\begin{split} 
	\left. g_{ij}\left(\partial - \bar{\partial}\right)\cc X^j+B_{ij}\left(\partial +\bar{\partial}\right)\cc X^j\right|_{\partial \Sigma}=0\,,
\end{split}
\end{equation}
where $\partial = \partial/\partial z$ and $\bar{\partial}=\partial/\partial \bar{z}$ {and $\partial \Sigma$ is mapped to the real line}. The propagator with these boundary conditions is  \cite{Fradkin:1985qd,Abouelsaood:1986gd,Callan:1986bc}
\begin{equation}
\begin{split} 
	\bigl\langle\cc X^i(z) \cc X^j (z') \bigr\rangle=& -\alpha^{\prime} \left(g^{ij}\log\left|z-z'\right|-g^{ij} \log\left|z-\bar{z}'\right|\right.\\
	&\hspace{70pt}+\left.G^{ij}\log\left|z-\bar{z}'\right|^2+\theta^{ij}\log\frac{z-\bar{z}'}{\bar{z}-z'}+D^{ij}\right),
\end{split} \label{eq:DDProp}
\end{equation}
where
\eq{
\arraycolsep2pt
\begin{array}{lclcl}
G_{ij}&=& \multicolumn{3}{l}{\displaystyle  \left[ g- B\op g^{-1} B\right]_{ij},}
\\[10pt]
G^{ij}&=& \displaystyle \left[(g+ B)^{-1}\right]^{ij}_S &=& \displaystyle +\left[(g+ B)^{-1}\op g\op(g- B)^{-1}\right]^{ij},\\
\theta^{ij}&=& \displaystyle  \left[(g+ B)^{-1}\right]^{ij}_A &=& \displaystyle 
- \left[(g+ B)^{-1}\op B\op(g- B)^{-1}\right]^{ij}. 
\end{array}
\label{SWMapping}
}
Here, $\left(\right)_{A}$ and $\left(\right)_{S}$ denote the antisymmetric and the symmetric part of the matrix. The constants $D^{ij}$ 
in \eqref{eq:DDProp} don't play an essential role for the subsequent analysis because they don't depend on $z$ or $z'$.

For the purpose of this review, we are only interested in the behaviour of the endpoints of the open string, which we will therefore focus on.
At the endpoints of the string we have $z = \tau$ and $z' = {\tau'}$.
As reviewed in detail in appendix~\ref{app_sw1}, the two-point function  \eqref{eq:DDProp} then becomes
\eq{
	\bigl\langle\cc X^i\left(\tau\right) \cc X^j\left(\tau'\right)\bigr\rangle=-\alpha^{\prime}G^{ij}\log\left(\tau-\tau'\right)^2+i \alpha' \pi
	\theta^{ij}\epsilon\left(\tau-\tau'\right)\,. \label{SW2p}
}
The object $G^{ij}$ plays the role of the effective metric for the open string.
Using \eqref{SW2p}, we can {now directly} compute 
the commutator for the endpoints:
\begin{equation}
\begin{split} 
	\left[\cc X^i(\tau),\cc X^j(\tau )\right]=T\left(\cc X^i\left(\tau\right)\cc X^j\left(\tau^-\right)-\cc X^i\left(\tau\right)\cc X^j\left(\tau^+\right)\right)=2 \pi \alpha' i \op \theta^{ij}\,.\label{SWcom}
\end{split}
\end{equation}
{This result can be summarised by mapping the description with a metric, $g$ and $B$-field to a different metric, $G$, and non-commutativity parameter, $\theta$, without a $B$-field, as in}
\begin{equation}\label{mapping}
\begin{split} 
	G+\theta=\left(g+ B\right)^{-1}\,.
\end{split}
\end{equation}

Here we reviewed the case of flat space and with constant $B$-field. {\cite{Cornalba:2001sm, Herbst:2001ai} showed} that this result can be generalized to non-constant B-field and curved background and that the mapping in the above equation \eqref{mapping} is also valid for non-constant backgrounds. It is also important to notice, that the associated non-commutative star-product 
\begin{equation}
\begin{split} 
	f(X)* g(X)=f\, g + \pi \alpha' i \op \theta^{ij}\partial_i f \partial_j g +\mathcal{O}\left(\theta^2\right)\,,
\end{split}
\end{equation}
being associative for constant $\theta$, 
does not need to have a Poisson structure. Therefore, we can simply use the formulae \eqref{SWMapping} to compute the open-string parameters. Because we just have a Kontsevich type of product, 
it is possible that we end up with non-zero Jacobiator as
shown in \cite{Cornalba:2001sm, Herbst:2001ai}.


\subsection{Commutators for T-dual open-string coordinates}
In this section, we compute the non-commutative relations for the mutually T-dual open-string  coordinates. Later, we are going to need these commutators for the non-associativity of the open string. 
So let us start to compute the commutator for the open-string coordinate $\cc X(z,\bar z)$ and the dual open-string coordinate $\tilde {\cc X}(z,\bar z)$. The free open-string field $\cc X(z,\bar z)\equiv \cc X_{\rm NN}(z,\bar z)$ with Neumann-Neumann boundary conditions  possesses the following mode expansion 
\begin{eqnarray}
 \cc X(z,\bar z)=
 q-i\alpha' p\log |z|^2+i\sqrt{\frac{\alpha'}{2}}\sum_{n\neq 0}\frac{1}{n}\alpha_n(z^{-n}+\bar z^{-n})
\, ,
\end{eqnarray}
where $q$ and $p$ are the center-of-mass position and momentum and where we suppressed the space-time index.
For the dual coordinate $\tilde {\cc X}(z,\bar z)\equiv \cc X_{\rm DD}(z,\bar z)$ with Dirichlet-Dirichlet boundary conditions we have
\begin{eqnarray}
 \tilde {\cc X}(z,\bar z)=
q_0+ \frac{1}{2\pi i}(q_1-q_0)\log \biggl({z\over \bar z}\biggr)+i\sqrt{\frac{\alpha'}{2}}\sum_{n\neq 0}\frac{1}{n}\alpha_n(z^{-n}-\bar z^{-n})
\, ,
\end{eqnarray}
with $q_0$ and $q_1$ denoting the start- and end-point of the open string. 
The two-point function between $\cc X(z,\bar z)$ and $\tilde {\cc X}(w,\bar w)$ can be deduced from the mode expansions above as
\begin{eqnarray}\label{OpenfreefieldcorA}
 \langle \cc X(z,\bar z)   \tilde  {\cc X}(w,\bar w)\rangle=
 -{\alpha'\over 2}\Biggl(\log\biggl({z-w\over \bar z-\bar w}\biggr)+\log\biggl({\bar z-w\over z-\bar w}\biggr)\Biggr)
 \, .
\end{eqnarray}
Comparing  this correlation function with the two-point function in \eqref{eq:DDProp}
we can immediately compute the equal time commutator between $\cc X$ and $\tilde {\cc X}$ at the end points of the open string:
\begin{equation}
\begin{split} 
	\left[\cc X(\tau),\tilde{\cc  X}(\tau )\right]=\pi \alpha' i\,.
\end{split}
\end{equation}
Unlike for closed strings, the commutator between  the open-string coordinate and its dual is non-zero, and in this sense they are indeed canonically conjugate to each other.
In the appendices
we will  compute  the equal time commutator among the open string coordinate $\cc X(z,\bar z)$ and the dual coordinate $\tilde {\cc X}(z,\bar z)$ in an alternative way as was done before for the closed 
string \cite{Bakas:2015gia}. 
In addition the analogous computation for closed strings is presented in the appendix.


\section{Open strings on a three-torus with (non)-geometric fluxes}

In this section we illustrate the non-associative behaviour of open-string coordinates in a 
non-geometric flux background. 
We apply a chain of T-duality transformations to a three-torus with $H$-flux, which leads to 
backgrounds with geometric $F$-flux, non-geometric $Q$-flux and 
non-geometric $R$-flux \cite{Dasgupta:1999ss,Kachru:2002sk,Hull:2004in,Shelton:2005cf}
\eq{
  \label{chain_t_duality}
  H_{ijk} \quad\xrightarrow{\hspace{10pt}T_k\hspace{10pt}}\quad
  F_{ij}{}^k \quad\xrightarrow{\hspace{10pt}T_j\hspace{10pt}}\quad
  Q_{i}{}^{jk} \quad\xrightarrow{\hspace{10pt}T_i\hspace{10pt}}\quad
  R^{ijk} \,.
}  
Here $T_i$ denotes a T-duality transformation along the direction $X^i$ of the $\mathbb T^3$
and the indices take values $i,j,k=1,2,3$. 
Note that a three-torus with one of the above choices of flux is only a toy example, and that this three-dimensional setting has to be embedded into a proper string-theory construction which usually requires the introduction of additional fluxes. For clarity we focus here only on one type of flux present in each duality frame, but more complicated settings are possible (see for instance \cite{Plauschinn:2018wbo}).

The open-string sector is characterized by a choice of D-branes, and we start from a
point-like D0-brane on the $H$-flux background. Performing a T-duality transformation perpendicular
to a D$p$-brane results in a D$(p+1)$-brane, and thus the $R$-flux background in \eqref{chain_t_duality} 
contains a $\mathbb T^3$-filling D3-brane.

We also remark that for the computation of commutators 
we use the mode expansion of  open-string coordinates in a flat background. 
This is a good approximation as long as the flux density is small, which can be achieved by a suitable choice of 
radii for the three-torus. For a proper string-background one has to check whether this can be achieved dynamically,
however, this is beyond the scope of the present paper. 
String-compactifications on nontrivially-fibered tori in this context have been studied  in a number of papers, 
for instance in \cite{Blumenhagen:2011ph,Andriot:2012vb,Lust:2017jox}, and evidence for the validity of this approach 
has been found.


\subsection{$\mathbb T^3$ with H-flux and D0-branes}

We consider a three-torus $\mathbb T^3$ with non-vanishing $H$-flux. The flux is quantized 
and we denote its non-trivial component by $H_{123} = H = N \in \frac{1}{\ell_{\rm s}}\mathbb Z$ 
where \raisebox{0pt}[0pt][0pt]{$\ell_{\rm s} = 2\pi\sqrt{\alpha'}$} denotes the string length.
The metric and Kalb-Ramond
$B$-field for this background can be specified by 
\eq{
\label{bb_001}
ds^2=& r_1^2\bigl(dX^1\bigr)^2+r_2^2\bigl(dX^2\bigr)^2 +r_3^2\bigl(dX^3\bigr)^2 \,, \\[2pt]
B_{12}=&H\op X^3 \,,
}
and we place a single point-like D0-brane into this background. Note that the  Kalb-Ramond field 
is not constant and depends on the closed string coordinate $\op X^3$.
The corresponding open strings 
therefore have Dirichlet boundary conditions along the directions of the three-torus. 
One can then show that the open-string coordinates with Dirichlet-Dirichlet (DD) boundary 
conditions commute, that is 
\eq{
\bigl[\tilde{\cc  X}_i,\tilde {\cc X}_j\bigr] \equiv \bigl[{\cc X}_{i\op}{}_{ {\rm  DD}},{\cc X}_{j\op}{}_{{\rm  DD}}\bigr]= 0 \,, \hspace{50pt} i,j=1,2,3 \,.
}
Let us also mention that one can introduce also D1- and D2-branes into this space 
(see for instance \cite{Cordonier-Tello:2018zdw}), while 
D3-branes are excluded due to the Freed-Witten anomaly cancellation condition \cite{Freed:1999vc}.
In the case of D2-branes, the resulting gauge theory on the brane will in general be non-commutative.


\subsection{Twisted torus with D1-branes}

We now perform a T-duality transformation of the background \eqref{bb_001}
along the $X^1$-direction. Using for instance the standard Buscher rules, we 
obtain a twisted torus with vanishing $H$-flux specified by
\eq{
\label{bb_002}
  ds^2&= \frac{1}{r_1^2} \bigl(d{X}^1-F\op X^3 dX^2 \bigr)^2+r_2^2\bigl(dX^2\bigr)^2 +r_3^2\bigl(dX^3\bigr)^2\,,\\
  B_{ij}&=0\,,
}
where $F=N$ is called the geometric flux.
Due to the T-duality transformation,  the boundary condition of the open string  along the $X^1$-direction changes 
from Dirichlet-Dirichlet  to Neumann-Neumann (NN). We therefore obtain a D1-brane 
with an  angle with respect to one side of the torus, where this angle is determined by the geometric flux parameter $F$.
(For a review of this mechanism see for instance \cite{Plauschinn:2018wbo}.)
The open-string coordinates are commutative in all three directions of the twisted torus
\eq{
  \arraycolsep2pt
  \begin{array}{lclcl}
  \displaystyle \bigl[\cc X^1,\tilde{\cc  X}_i\bigr] &\equiv& \displaystyle   \bigl[\cc X^1_{\rm NN},{\cc X}_{i\op{\rm DD}}\bigr]
  &=& 0 \,,
  \\[10pt]
  \displaystyle \bigl[\tilde{\cc  X}_2,\tilde {\cc X}_3\bigr] &\equiv&  \bigl[{\cc X}_{2\op{\rm DD}},{\cc X}_{3\op{\rm DD}}\bigr] &=& 0 \,, 
  \end{array}
  \hspace{60pt} i=2,3 \,.
}


\subsection{Q-flux background with D2-branes} 

As a next step, we perform a T-duality transformation along the  $X^2$-direction of the twisted-torus background \eqref{bb_002}. 
The resulting configuration is a {so-called} T-fold \cite{Hull:2004in,Dabholkar:2005ve,Hull:2006va,Hull:2009sg},
for which the metric and  non-trivial Kalb-Ramond $B$-field component take the following form
\eq{
  ds^2=& \frac{r_2^2\bigl(d{X}^1\bigr)^2+r_1^2\bigl(d{X}^2\bigr)^2}{r_1^2 r_2^2+ \bigr(Q\op X^3\bigl)^2}+r_3^2\bigl(dX^3\bigr)\,,\\
  B_{12}=& \frac{Q\op X^3}{r_1^2 r_2^2+ \bigr(Q\op X^3\bigl)^2}\,. \label{Qflux}
}
The parameter $Q=N$ is called the non-geometric $Q$-flux.
Note that under $X^3\to X^3+\ell_{\rm s}$ this metric and $B$-field are not globally-defined using 
diffeomorphisms and gauge transformations, but are consistent as string-theory backgrounds
when using T-duality transformations as transition functions. 
Furthermore, under T-duality the boundary conditions along the $X^2$-direction change from DD to NN, and we have 
a D2-brane along the directions $X^1$ and $ X^2$.

As we have reviewed above, for a D2-brane in a non-trivial $B$-field background we expect the 
corresponding endpoints of open strings to be non-com\-mu\-tative. 
This can be seen by applying the Seiberg-Witten map \cite{Seiberg:1999vs} to the above configuration. 
More concretely, the metric and bi-vector in the open-string frame \eqref{eq:DDProp} for the T-fold \eqref{Qflux}
are obtained using \eqref{mapping} as
\eq{
d\hat s^2=& r_1^2\bigl(dX^1\bigr)^2+r_2^2\bigl(dX^2\bigr)^2 +\frac{1}{r_3^2}\bigl(dX^3\bigr)^2 \,, \\[2pt]
\theta^{12}=&-Q\op X^3 \,,
}
where $\theta^{12}$ is equal to the bi-vector $\beta^{ij}$ known from 
generalized geometry \cite{Hitchin:2004ut,Hitchin:2005in}
and double field theory \cite{Hull:2009mi,Hohm:2010pp,Hohm:2012mf,Hohm:2013bwa}.
This bi-vector gives rise to the 
$Q$-flux as \cite{Andriot:2011uh,Andriot:2012wx,Andriot:2012an,Blumenhagen:2012nk,Blumenhagen:2012nt,Blumenhagen:2013aia}
\begin{equation}
Q_k{}^{ij}=\partial_k\beta^{ij}\, .
\end{equation}
Now, as already discussed in \cite{Grange:2006es}, a non-zero $\theta^{ij}$ leads to a non-commutative behaviour of the 
open-string coordinates
\begin{equation}\label{qfluxcom}
  \bigl[\cc X^1,\cc X^2\bigr]\equiv \bigl[\cc X^1_{\rm NN},\cc X^2_{\rm NN}\bigr]=2 \pi \alpha' i \op \theta^{12}= -2\pi\alpha' i \op Q X^3\,,
\end{equation}
in particular, the $Q$-flux controls this commutator.
However note that whereas the corresponding expression in the closed-string case was determined by the winding 
number $\tilde p^3$ (cf. equation\eqref{rel_001}), now the coordinate $X^3$ appears
on the r.h.s. of the commutator.

We also mention that the result \eqref{qfluxcom} can also be obtained through a direct computation 
as in \cite{Chu:1998qz}. Indeed, using the mode expansion of open-string coordinates 
with NN boundary conditions the commutator can be evaluated explicitly. 
Let us then recall from  \cite{Chu:1998qz} the equal-time commutator of two 
open-string coordinates with NN boundary conditions  as
\eq{
  \label{qfluxcomm_2}
  \bigl[ \cc X_{\rm NN}^i(\tau,\sigma), \cc X^j_{\rm NN} (\tau,\sigma') \bigr] = 2\pi\alpha' \op i\,\Omega(\sigma,\sigma')
  \Bigl[ \bigl( g - B\op g^{-1} \op B \bigr)^{-1} B\op g^{-1} \Bigr]^{ij}\,, 
}
where $\Omega(\sigma,\sigma')$ is $+1$ for $\sigma=\sigma'=0$,
$-1$ for $\sigma=\sigma'=\pi$ and zero for $\sigma\neq\sigma'$. 
Using then the explicit expressions \eqref{Qflux} for the metric and $B$-field of the T-fold, 
we obtain \eqref{qfluxcom} at the endpoints of the open string.

Let us  briefly comment on the appearance of the closed-string coordinate  in \eqref{qfluxcom}: 
we can interpret the T-fold background as a non-geometric $\mathbb T^2$-fibration over a circle. 
Computing the commutator \eqref{qfluxcom} fiberwise at a fixed point $X^3$ on the base, 
the non-commutativity parameter $\theta^{12}$ is constant on the fiber and we obtain the well-known result. 
In a second step we extend the fiber-wise result along the base-cycle, which means that 
$X^3$ can now vary. Since $\theta^{12}$ depends on the closed-string coordinate $X^3$, also the 
commutator \eqref{qfluxcom} depends on the closed-string $X^3$. 
This conclusion also agrees with \eqref{qfluxcomm_2}, in which the closed-string metric and Kalb-Ramond field 
appear on the right-hand side. 
It would be desirable to replace the fiber-wise computation of the commutator 
by commutators between interacting fields as it was done for instance in \cite{Blumenhagen:2011ph}, however, this is beyond 
the scope of this work.


\subsection{R-flux background with D3-branes} 

As last step, we perform a T-duality transformation along the $X^3$-direction on the $Q$-flux background
\eqref{Qflux}. Since $X^3$ is not a direction of isometry, the Buscher rules cannot be applied and the 
duality transformation can only be done formally. 
On general grounds, and in agreement with the picture \eqref{chain_t_duality}, we expect that 
the $Q$-flux is transformed into an $R$-flux.\footnote{
A convenient \emph{space-time} framework for studying non-geometric fluxes is double field theory
\cite{Hull:2009mi,Hohm:2010pp,Hohm:2012mf,Hohm:2013bwa}. 
For the computation of open-string commutators we are however interested in a \emph{world-sheet} theory,
which is provided by string theory with the inclusion of winding coordinates $\tilde X_i$.} 
Denoting by $\tilde X_3$ the coordinate dual to the closed-string 
coordinate $X^3$, the 
T-dual bi-vector is expected to be of the form
\begin{equation}
	\theta^{12}  =  \beta^{12} = - R \tilde{X}_3 \,,
\end{equation}
with $R=N$, 
and the resulting $R$-flux is computed 
via the equation 
 \cite{Andriot:2011uh,Andriot:2012wx,Andriot:2012an,Blumenhagen:2012nk,Blumenhagen:2012nt,Blumenhagen:2013aia}
\begin{equation}
R^{ijk}=3\op \tilde\partial^{[i}\beta^{jk]}\, .
\end{equation}
Furthermore, the D2-brane on the $Q$-flux background is mapped to a D3-brane on 
the $R$-flux background.
Consequently, performing the T-duality transformation $X^3\leftrightarrow \tilde X_3$ in eq.~\eqref{qfluxcom},
 the open-string commutation relations between the two coordinates $X^1$ and $X^2$ in the $R$-flux frame are given as
\begin{equation}\label{rfluxcom}
 \bigl[\cc X^1,\cc X^2\bigr]\equiv\bigl[\cc X^1_{\rm NN},\cc X^2_{\rm NN}\bigr]
=-2 \pi\alpha' i \op R \op\tilde X_3\,,
\end{equation}
with  $\tilde X_3$ the dual closed-string coordinate of the background space.

We now want to use the result shown in \eqref{rfluxcom} to evaluate a three-bracket for the 
$R$-flux background. More concretely, similar to \eqref{asso} we define a three-bracket as
\eq{
  \label{assos_open}
    \Bigl[ \cc X^1,\cc X^2, \cc X^3 \Bigr]\equiv
  \Bigl[ \cc X^1_{\rm NN},\cc X^2_{\rm NN}, \cc X^3_{\rm NN} \Bigr] = 
  \Bigl[ \bigl[ \cc X^1_{\rm NN},\cc X^2_{\rm NN} \bigr], \cc X^3_{\rm NN} \Bigr] + \mbox{cyclic.}
}  
Next, we compute the commutator between  an open-string coordinate and 
a dual closed-string coordinate.  This is done in appendix~\ref{app_oc_com}, in particular in equation \eqref{CommNNdclosed}
we obtain
\begin{equation}\label{bb_007}
\bigl[\cc X^3_{\rm NN},\tilde X_3\bigr]
= \pi \alpha' \op i \, .
\end{equation}
Then we find for the open-string associator \eqref{assos_open}
\begin{equation}\label{r3bracket}
	\bigl[ \cc X^1_{\rm NN},\cc X^2_{\rm NN},\cc X^3_{\rm NN}\bigr]
	=- 2 \pi \alpha' i\op R \op\bigl[\tilde{X}_3,\cc X^3_{\rm NN}\bigr]=  - 2 \pi^2 \alpha'^2  R  \,.
\end{equation}
Thus, we derive in this case a non-associative behaviour of the open-string position coordinates. 
All three fields $\cc X^1$, $\cc X^2$ and $\cc X^3$ in the three-bracket eq.~\eqref{r3bracket}  correspond to open-string coordinates with end points along the D3-branes, 
i.e. to open-string coordinates with Neumann-Neumann boundary conditions.
The effective open-string gauge theory on the D3-branes is expected to be a non-associative gauge theory, which possibly possesses an $L_\infty$ structure along the lines discussed in  \cite{Hohm:2017cey,Blumenhagen:2018kwq}.


\subsection{Decoupling limit}

In order to have a gauge theory on the D-brane world-volumes, we have to decouple the closed string from the open strings, i.e we must consider the limit where gravity decouples from the D-branes, and show that the non-associative algebra survives. This is the limit of infinite string tension, i.e. the limit where 
\begin{equation}
\alpha' \rightarrow 0\,.
\end{equation}
To investigate this limit, we will determine how certain operators and parameters scale as $\alpha' \rightarrow 0$. We choose the conventions in which the open and closed string coordinates $X^i$ are dimensionless, i.e. stay constant in this limit.
It follows that the metric $g_{ij}$  and the B-field $B_{ij}$ have dimension $\alpha'$.\footnote{In this section we 
change our conventions in order to be in line with the existing literature.}
As a first check, we see that the standard Seiberg/Witten non-commutativity parameter $\theta^{ij}$ in \eqref{openbasic} behaves as
\begin{equation}
\theta^{ij} \sim (\alpha')^{-1}\,,
\end{equation}
and the open string commutator in this equation stays constant in the decoupling limit:
\begin{equation}
\begin{split} 
	\left[ \cc X^i, \cc X^j\right]_{\rm open}=2\pi \alpha' i \op\theta^{ij}
\sim const\,.
\end{split}
\end{equation}

Next, let us determine the scaling behaviour of the dual coordinates.
Since $\tilde X_i = (g+B)_{ij} X_L^j-(g-B)_{ij}  X_R^j$  and since $X_{L/R}^i$ are dimensionless, it the follows that the dual coordinates $\tilde{X}_i$ have dimension $\alpha'$, i.e. they scale as
\begin{equation}
\tilde X_i \sim \alpha'\,.
\end{equation}
This is also consistent with the open string commutator
\begin{equation}
\begin{split} 
	\left[\cc X^i,\tilde{\cc  X}_j \right]=\pi \alpha' i\, \delta^i_j\,.
\end{split}
\end{equation}
The scaling behaviour of the dual string coordinates can be also derived in the following way.
Namely the dual momentum on a circle is given as
\begin{equation}\tilde p^i={r^i\over \alpha'} \sim   ( \alpha')^{-1}   \, .
\end{equation}
Using the commutation relation $\left[\tilde p^i,\tilde{  X}_j\right]=\delta^i_j$, one also obtains that $\tilde X_i \sim \alpha'$.
So we see that in the zero-slope limit the dual coordinates are vanishing, which is just a manifestation of the section condition in double field theory.

Now let us determine how the non-geometric fluxes behave in the decoupling limit.
Since $(g+B)^{-1} = G^{-1} + \beta$, we find that  
\begin{equation}
\beta \sim \alpha'^{-1}\,,
\end{equation}
which leads for the $Q$-flux:
\begin{equation}
Q_k{}^{ij}=\partial_k\beta^{ij} \sim
(\alpha')^{-1}\, .
\end{equation}
For the R-flux we find
\begin{equation}
R^{ijk}=\tilde\partial^{[k}\beta^{ij]}
\sim (\alpha')^{-2}\,.
\end{equation}
Having derived these relations, we can finally determine how the commutation relations behave in the decoupling limit.
For the D2-branes in the $Q$-flux background we get that
\begin{equation}
  \bigl[\cc X^1,\cc X^2\bigr]=2 \pi \alpha' i \op \theta^{12}= -2\pi\alpha' i \op Q X^3
  \sim{\rm const}
   \,.
\end{equation}
This shows that in the decoupling limit the open string coordinates on the D2-branes in the $Q$-flux background do not commute with each other.
For the D3-branes in the $R$-flux background we get that
\begin{equation}
 \bigl[\cc X^1,\cc X^2\bigr]
=-2 \pi\alpha' i \op R \op\tilde X_3
\sim {\rm const}\,.
\end{equation}
Hence in the decoupling limit the open string coordinates
on the D3-branes in the $R$-flux background
 do not commute with each other.
Let us finally see what happens with the non-associativity, namely with the 3-bracket  of the open string coordinates on the D3-branes in the $R$-flux background:
\begin{equation}
	\bigl[ \cc X^1,\cc X^2,\cc X^3\bigr]
	=  - 2 \pi^2 \alpha'^2  R \sim {\rm const}\,.
\end{equation}
Therefore also the non-associativity survives in the decoupling limit of gravity, which means there indeed should exist a non-associative
gauge theory on the D3-brane world-volume in the $R$-flux background.


\section{Summary}

Let us now summarize the results of the previous section for open-string 
commutators  in $Q$- and $R$-flux backgrounds, and compare these
to the closed-string case. 
Denoting by
$\tilde X_i$ again the coordinates dual to the usual closed-string coordinates $X^i$,
and by
$p_i$ and $\tilde p^i$ the momentum and dual winding numbers
for the closed string, we have the following structure: 
\eq{
  \renewcommand{\arraystretch}{1.8}
  \arraycolsep15pt
  \begin{array}{c||l@{\op\op}c@{\op\op}l|l@{\op\op}c@{\op\op}l}
  & \multicolumn{3}{c|}{\mbox{closed string}} & \multicolumn{3}{c}{\mbox{open string}}
  \\
  \hline\hline
  \mbox{$Q$-flux} & \bigl[X^i,X^j\bigr]&=&i\op  Q_k{}^{ij}\op\tilde p^k & 
  \bigl[\cc X^i,\cc X^j\bigr]&=&i\op  Q_k{}^{ij}\op X^k
  \\
  \hline
  \multirow{3}{*}{\mbox{$R$-flux}} & \bigl[X^i,X^j\bigr]&=&i\op  R^{ijk}\op p_k & 
  \bigl[\cc X^i,\cc X^j\bigr]&=&i\op  R^{ijk}\op \tilde X_k 
  \\
  &  \bigl[X^i,p_j\bigr] &=& i \op \delta^i{}_j & 
   \bigl[\cc X^i,\tilde {\cc X}_j\bigr]&=& i \op \delta^i{}_j 
  \\
  &  \multicolumn{3}{l|}{\bigl[X^i,X^j, X^k\bigr]=  R^{ijk}} &
  \multicolumn{3}{c}{\bigl[\cc X^i,\cc X^j,\cc  X^k\bigr]=  R^{ijk}}
  \end{array}
}
where we omitted all numerical factors. 
For open strings in a $Q$-flux background the commutator between the coordinates defines 
a Lie-algebra-valued non-commutative algebra, since the non-commu\-tativity is linear in the coordinates with Lie-algebra valued structure constants $Q_k{}^{ij}$. 
The open string non-commu\-tative gauge theory on the D2-branes possesses an underlying
 $*$-pro\-duct, which is now coordinate dependent:
\begin{equation}
\begin{split} 
	f(X)* g(X)=f\, g + \frac{i}{2}    Q_k{}^{ij}\op X^k        \partial_i f \partial_j g +\mathcal{O}\bigl(Q^2\bigr)\,.
\end{split}
\end{equation}
For open strings in an $R$-flux background it is natural to expect that the gauge theory on the D3-branes will be a non-associative gauge theory.
Now the open-string  $*$-product for the $R$-flux case takes the form
\begin{equation}\label{star_ppp}
\begin{split} 
	f(X)* g(X)=f\, g + \frac{i}{2}    R^{ijk}\op \tilde X_k        \partial_i f \partial_j g +\mathcal{O}\bigl(R^2\bigr)\,,
\end{split}
\end{equation}
and one also obtains a related non-associative $\triangle$-product of the following form \cite{Blumenhagen:2011ph,Mylonas:2012pg,Bakas:2013jwa}: 
\begin{equation}
\begin{split} 
	f(X)\triangle g(X)\triangle h(X)=f\, g\, h + \frac{1}{6}    R^{ijk}        \partial_i f \partial_j g        \partial_k h  +\mathcal{O}\bigl(R^2\bigr)\,.
\end{split}
\end{equation}
An interesting question is whether there exists an analogue of the Seiberg-Witten map, which maps a 
non-associative gauge theory with product \eqref{star_ppp} to a commutative one.
The result obtained above provides the building block for constructing such a non-associative
gauge theory, and a good starting point for determining the Seiberg-Witten map is \cite{Blumenhagen:2018shf}.
We leave this questions as well as the study of properties of the non-associative gauge theory, 
for future work. 

\bigskip
Concerning the non-associativity of the D3-branes on an $R$-flux background,
the authors of \cite{Hull:2019iuy} come to a conclusion different from ours.
Specifically, in \cite{Hull:2019iuy}  it is argued that the $*$-product defined in  
\eqref{star_ppp} is associative, whereas 
in our case this product leads to a non-associative structure due to the non-vanishing 
commutator between $\mathsf X$ and $\tilde{\mathsf X}$ shown in \eqref{bb_007}.
The main conceptual difference between the work in \cite{Hull:2019iuy}  and ours is that the former performs an analysis mostly at the level of the effective space-time theory (DFT and aspects of doubled geometry) while we work with a two-dimensional world-sheet description and compute commutators therein. Both approaches focus on different aspects of the problem, but should be equally valid. However, in DFT a non-vanishing associator violates the strong constraints and hence 
can not be seen. It is important to resolve the tension between the results in \cite{Hull:2019iuy} and ours, and 
work in this direction is under way.


\vskip0.5cm

\subsection*{Acknowledgements}
We thank R. Blumenhagen, L. Freidel, C. Hull and R. Szabo for very useful discussions.
This work was partially supported by the ERC Advanced Grant ``Strings and Gravity'' 
(Grant No. 320045) {and the ERC Advanced Grant ``Exceptional Quantum Gravity'' (Grant No. 740209)}.


\appendix


\section{Commutators between string coordinates and dual string coordinates}


\subsection{Closed string}

Let us start  with a review regarding the commutation relations for the  closed string (see \cite{Bakas:2015gia}). 
(For this appendix we set $\alpha' = 1$.)
Time ordering becomes radial ordering in the complex $z$-plane,  and hence the equal time commutator between the closed string coordinate $X(\tau,\sigma)$ and the dual closed string coordinate $\tilde X(\tau,\sigma)$ is given as: 
\eq{
&\lbrack \tilde X(\tau,\sigma),X(\tau,\sigma')\rbrack=\lbrack \tilde X(z,\bar z),X(w,\bar w)\rbrack_{|z|=|w|}=
 \\ 
&\hspace{40pt}= \lim_{\delta\rightarrow 0_+}\biggl[\tilde X(z,\bar z)X(w,\bar w)\Bigr\rvert_{|z|=|w|+\delta}-X(w,\bar w) \tilde X(z,\bar z)\Bigr\rvert_{|z|=|w|-\delta}\biggr]\,,
}
where we suppressed the space-time index.
The two point function between $X$ and $\tilde X$ is
\begin{eqnarray}\label{freefieldcorA}
 \langle X(z,\bar z)         \tilde X(w,\bar w)\rangle=-{1 \over 2}\log{z-w\over \bar z-\bar w}\, ,
\end{eqnarray}
as it can be computed directly from the mode expansion of the left and the right movers.
So we obtain for the commutator
\eq{
& \lbrack \tilde X(z,\bar z),X(w,\bar w)\rbrack_{|z|=|w|}
\\
&\hspace{60pt}=
-{1 \over 2} \lim_{\delta\rightarrow 0_+}\biggl[\log{z-w\over \bar z-\bar w}\biggr\rvert_{|z|=|w|+\delta}
-\log{z-w\over \bar z-\bar w}\biggr\rvert_{|z|=|w|-\delta}
\biggr]\,.
}
Choosing $\tau=\pm\delta$, $\tau'=0$ 
implies $z=e^{\pm\delta-i\sigma}$, $w=e^{-i\sigma'}$, 
and with $\varphi=\sigma-\sigma'$
the above expression becomes
\eq{
\label{closedcom}
\lbrack \tilde X(z,\bar z),X(w,\bar w)\rbrack_{|z|=|w|}
-{1 \over 2} \lim_{\delta\rightarrow 0_+}\biggl[\log{e^\delta e^{-i\varphi}-1\over e^\delta e^{i\varphi}-1}-\log{e^{-\delta} e^{-i\varphi}-1\over e^{-\delta} e^{i\varphi}-1}\biggr]\,.
}
As described in \cite{Bakas:2015gia}, the commutator between coordinate and dual coordinate can then be evaluated as:
\eq{
\lbrack \tilde X(\tau,\sigma),X(\tau,\sigma')\rbrack 
=\lbrack \tilde X(z,\bar z),X(w,\bar w)\rbrack_{|z|=|w|}
=\pi i\op \epsilon(\varphi)
=\pi i\op \epsilon(\sigma-\sigma')\, ,
}
where $\epsilon(\sigma-\sigma')$ denotes the step function.
We can set $\sigma=\sigma'$ and the commutator is vanishing:
\begin{equation}\label{commphi}
\lbrack \tilde X(\tau,\sigma),X(\tau,\sigma)\rbrack=\pi i\op\epsilon(\varphi)=\pi i\op\epsilon(0)=0\, .
\end{equation}

Actually,  a more precise procedure to get the space time interpretation of this commutator is to consider the integrated version of the relevant two-di\-men\-sio\-nal operators.
Let us back up and  demonstrate this quickly first for the canonical commutator between coordinate and momentum operator. Here we have
\begin{equation}
\lbrack  X(\tau,\sigma),P(\tau,\sigma')\rbrack=\lbrack  X(\tau,\sigma),\partial_\tau X(\tau,\sigma')\rbrack=
2\pi i\op \delta(\varphi)=2\pi i\op\delta(\sigma-\sigma')\, .
\end{equation}
Next we consider the equal time commutator of the operators integrated over the string:
\eq{
\lbrack X(\tau),P(\tau)\rbrack &= \left[  \frac{1}{2\pi}\int_0^{2\pi}d\sigma X(\tau,\sigma),
\frac{1}{2\pi}\int_0^{2\pi}d\sigma' P(\tau,\sigma')\right]
\\
&= \frac{ i}{2\pi} \int_0^{2\pi} d\sigma \int_0^{2\pi} d\sigma' \,\delta(\sigma-\sigma')
\\[6pt]
&= i\, .
}
After this first exercise let us return to equal-time commutator between the closed-string coordinate and its dual, where we now also integrate over the string:
\eq{
\label{commexact}
\lbrack \tilde X(\tau), X(\tau)\rbrack&= 
\left[ \frac{1}{2\pi}\int_0^{2\pi}d\sigma \tilde X(\tau,\sigma),\frac{1}{2\pi}\int_0^{2\pi}d\sigma' X(\tau,\sigma')\right] 
\\
&=\frac{i}{4\pi} \int_0^{2\pi}d\sigma\int_0^{2\pi}d\sigma' \epsilon(\sigma-\sigma')
\\[6pt]
&=0 \, .
}
So we see that this precise method gives the same result as the short version in (\ref{commphi}).
In conclusion, we see that for closed strings the equal-time commutator between the string coordinate and the associated dual string coordinate is zero. Note that 
we derive a  conclusion, which is  different from other work in the literature \cite{Freidel:2017wst,Freidel:2017nhg} concerning the intrinsic non-commutativity of closed strings.
Specifically, the actual difference comes between the calculation of \cite{Freidel:2017wst,Freidel:2017nhg} and our calculation is the different treatment of the zero modes.


\subsection{Open string}

From the open-string correlation function (\ref{SW2p}) 
we now obtain the following  commutator:
\eq{
&\lbrack  \cc X(\tau,\sigma),\tilde {\cc X}(\tau,\sigma')\rbrack=\lbrack  \cc X(z,\bar z),\tilde{\cc  X}(w,\bar w)\rbrack_{|z|=|w|}
\\
&\hspace{40pt}= \lim_{\delta\rightarrow 0_+}\biggl[\cc X(z,\bar z)\tilde{\cc  X}(w,\bar w)\Bigr\rvert_{|z|=|w|+\delta}
-\tilde {\cc X}(w,\bar w) \cc  X(z,\bar z)\Bigr\rvert_{|z|=|w|-\delta}\biggr] \,.
}
This expression becomes
\eq{\label{app_rel_9309}
\lbrack \cc  X(z,\bar z), \tilde{\cc  X}(w,\bar w)\rbrack_{|z|=|w|} \hspace{40pt}&
\\
=-{1\over 2} \lim_{\delta\rightarrow 0_+}\Biggl\lbrack\quad
&\biggl(\log{z-w\over \bar z-\bar w}+\log{\bar z-w\over z-\bar w}\biggr)
_{|z|=|w|+\delta}
\\
-&\biggl(\log{z-w\over \bar z-\bar w}+\log{\bar z-w\over z-\bar w}\biggr)
_{|z|=|w|-\delta} \quad
\Biggr\rbrack\,.
}
Again using coordinates $z=e^{\pm\delta-i\sigma}$, $w=e^{-i\sigma'}$      and $\varphi=\sigma-\sigma'$,
equation \eqref{app_rel_9309}
becomes:
\eq{
\label{opencom}
\lbrack \cc  X(z,\bar z), \tilde {\cc X}(w,\bar w)\rbrack_{|z|=|w|}\hspace{20pt}&
\\
=-{1 \over 2} \lim_{\delta\rightarrow 0_+}\Biggl\lbrack
\quad&
\log{e^{\delta-i\sigma}-e^{-i\sigma'}\over  e^{\delta+i\sigma}-e^{i\sigma'}}+\log{e^{\delta+i\sigma}-e^{-i\sigma'}\over e^{\delta-i\sigma}-e^{i\sigma'}}
\\
-&
\log{e^{-\delta-i\sigma}-e^{-i\sigma'}\over e^{-\delta+i\sigma}-e^{i\sigma'}}-\log{e^{-\delta+i\sigma}-e^{-i\sigma'}\over e^{-\delta-i\sigma}-e^{i\sigma'}}
\quad\Biggr\rbrack
\\
=
-{1 \over 2} \lim_{\delta\rightarrow 0_+}\Biggl\lbrack
\quad &\log{e^{\delta-i\varphi}-1\over  e^{\delta+i\varphi}-1}+\log{e^{\delta+i\varphi+2i\sigma'}-1\over e^{\delta-i\varphi-2i\sigma'}-1}
\\
-&
\log{e^{-\delta-i\varphi}-1\over e^{-\delta+i\varphi}-1}-\log{e^{-\delta+i\varphi+2i\sigma'}-1\over e^{-\delta-i\varphi-2i\sigma'}-1}
\hspace{23.5pt}\Biggr\rbrack\,.
}
Let us compare the open-string commutator in eq.\eqref{opencom} with the closed-string commutator in eq.\eqref{closedcom}.
We see that the open-string commutator has two additional terms, which lead to the following extra term in the final result
\eq{
\lbrack \tilde {\cc X}(\tau,\sigma),\cc X(\tau,\sigma')\rbrack 
&=\pi i \Bigl[ \epsilon(\varphi)+\epsilon(\varphi+2\sigma')\Bigr]
\\
&=\pi i \Bigl[ \epsilon(\sigma-\sigma')+\epsilon(\sigma+\sigma') \Bigr]\, .
}
Setting  $\sigma=\sigma'$ we obtain:
\begin{equation}
\lbrack \tilde {\cc X}(\tau,\sigma),\cc X(\tau,\sigma)\rbrack
=\pi i\op \epsilon(2\sigma)\, ,
\end{equation}
which is non-zero for $\sigma>0$.
Actually, we can integrate this over the open string and the integrated result is 
\begin{equation}
\frac{1}{\pi} \int_0^{\pi} d\sigma \op \lbrack \tilde{\cc  X}(\tau,\sigma),\cc X(\tau,\sigma)\rbrack
= i\int_0^{\pi} d\sigma\,\epsilon(2\sigma)=\pi \op i\,.
\end{equation}
Again let us look at the equal time commutator of the operators integrated over the open string:
\eq{
\label{commopenexact}
\lbrack \tilde{\cc  X}(\tau), \cc X(\tau)\rbrack&=
\left\lbrack \frac{1}{\pi} \int_0^{\pi} d\sigma \tilde {\cc X}(\tau,\sigma),\frac{1}{\pi} \int_0^{\pi} d\sigma' \cc X(\tau,\sigma')\right\rbrack
\\
&=\frac{i}{\pi}\int_0^{\pi }d\sigma\int_0^{\pi}d\sigma' \bigl(\epsilon(\sigma-\sigma')+\epsilon(\sigma+\sigma')\bigr)
\\[6pt]
&=\pi\op i\, .
}
This result agrees with the previous equation, where we have applied a short-cut procedure.
In the next appendix we complement these open string CFT computations by some slightly different but equivalent methods
to derive the commutation relations.


\section{Open-string computations}

In this appendix we summarize some technical details of the computation 
of two-point functions and commutators for Neumann-Neumann (NN)
and Diri\-chlet-Diri\-chlet (DD) open strings.


\subsection{Two-point function I}
\label{app_sw1}

We start by reviewing the computation of Seiberg and Witten \cite{Seiberg:1999vs} for the two-point function 
of two NN open-string coordinates on the boundary. 
The open-string coordinates $\cc X_{\rm NN}^i(z,\bar z)$ are functions on the upper half-plane parametrized 
by $z \in\mathbb C$ with ${\rm Im}\op z\geq0$, and the boundary of the open-string world-sheet is 
given by ${\rm Im}\op z = 0$.
Their starting point is the open-string two-point function \cite{Fradkin:1985qd,Abouelsaood:1986gd,Callan:1986bc}
\eq{	
\label{app_comp_01}
\bigl\langle \cc X_{\rm NN}^i(z,\bar z)\cc X^j_{\rm NN}(z',\bar z')\bigr\rangle=
&-\alpha^{\prime}\biggl[\,  g^{\,ij}\log|z-z'|-g^{\,ij}\log|z-\bar{z}'|
\\
&\hspace{22pt}+G^{\,ij}\log|z-\bar{z}'|^2+\op\theta^{\,ij}\log\frac{z-\bar{z}'}{\bar{z}-z'}+D^{\,ij}\,\biggr]
\\
=& -\alpha^{\prime}\biggl[\, \frac{1}{2}\op g^{ij} \log\frac{1+y^2}{1+x^2}
+	G^{\,ij}  \log\left( 1+x^2\right)
\\
&\hspace{22pt}+G^{\,ij}\log\left(\tau - \tau'\right)^2+\op \theta^{\,ij}\log\frac{1+i x}{1-i x}+D^{\,ij}
	\,\biggr],
}
where the matrices $g^{\,ij}$, $G^{\,ij}$, $\theta^{\,ij}$ and $D^{\,ij}$ were introduced in \eqref{SWMapping}.
Note in particular that $D^{ij}$ does not depend on $z$ or $z'$. 
We also introduced complex coordinates $z = \tau + i\op\sigma$ and 
$z' = \tau' + i\op\sigma'$
on the world-sheet, and defined the combinations
\eq{
\label{rel_009}
x=\frac{\sigma+\sigma'}{\tau - \tau'}\,, \hspace{50pt}
y = \frac{\sigma-\sigma'}{\tau - \tau'}\,.
}
We are now interested in the short-distance behaviour of the two-point function \eqref{app_comp_01} evaluated at the 
boundary $\sigma = \sigma'=0$. 
There is an ambiguity in taking this limit,  and here we follow
\cite{Seiberg:1999vs} as
\eq{
\label{SWlimit}
\arraycolsep1.5pt
\begin{array}{lcl}
\sigma=\sigma' & \rightarrow& 0 \\[2pt]
\tau-\tau' &\rightarrow&  0 
\end{array}
\hspace{40pt}\mbox{such that}\hspace{40pt}
\begin{array}{ccl}
|x| &=& \text{fixed}\gg 1\,, \\[2pt]
y &=& 0 \,.
\end{array}
}
The constant $x$-dependent terms in  \eqref{app_comp_01} proportional to $g^{ij}$ and $G^{ij}$ 
are cancelled by a convenient choice of $D^{\,ij}$, 
while the $\theta^{ij}$-term for large $x$ gives a step function depending on the sign of $x$.
In particular, with
\eq{
  \epsilon(\tau - \tau') = \left\{\begin{array}{l@{\hspace{15pt}}l}
  +1 & \tau>\tau' \\[4pt]
  -1 & \tau<\tau'
  \end{array}\right.
}
the two-point function \eqref{app_comp_01} evaluated on the boundary $\partial \Sigma$ of the world-sheet
becomes
\eq{
\label{app_comp_02}
\bigl\langle \cc X_{\rm NN}^i(z,\bar z)\cc X^j_{\rm NN}(z',\bar z')\bigr\rangle_{\partial\Sigma}
=-\alpha^{\prime} G^{\,ij}\log\left(\tau - \tau'\right)^2+i \alpha'\pi \theta^{\,ij}\epsilon(\tau-\tau')	\,.
}

Given the two-point function on the boundary \eqref{app_comp_02}, for a free theory also the 
equal-time commutator of two open-string coordinates on the boundary 
can be determined. In particular, one has \cite{Seiberg:1999vs} 
\eq{
\left[ \cc X_{\rm NN}^i(\tau),\cc X^j_{\rm NN}(\tau)\right]
&= T\left(\cc X^i_{\rm NN}(\tau)\cc X^j_{\rm NN}(\tau^-)-
\cc X^i_{\rm NN}(\tau)\cc X^j_{\rm NN}(\tau^+)\right)
\\[4pt]
&= 2 \pi \alpha' i \op \theta^{\,ij}\,,
}
where $T(\ldots)$ denotes time ordering. Alternatively, one can perform 
a mode expansion of the fields $\cc X^i_{\rm NN}(z)$ and determine 
the commutator explicitly, which leads to the same result \cite{Chu:1998qz}.


\subsection{Two-point function II}

We are going to perform a similar computation for the two-point function
of open-string coordinates with NN and DD boundary conditions. 
We start by recalling the mode expansions of a single open-string coordinate as
\eq{
\label{app_modes}
\arraycolsep2pt
\begin{array}{l@{\hspace{30pt}}lclcl}
\rm NN &  \cc X_{\rm NN}(z,\bar z) &=&
\displaystyle  q-i\alpha' p\log |z|^2&+&\displaystyle i\sqrt{{\alpha' \over 2}}   \sum_{n\neq 0}{1\over n}\alpha_n(z^{-n}+\bar z^{-n}) \,,
\\
\rm DD & \cc X_{\rm DD}(z,\bar z) &=&\displaystyle
q_0+ \frac{q_1-q_0}{ 2\pi i}\log \frac{z}{\bar z}&+&\displaystyle i\sqrt{{\alpha' \over 2}}\sum_{n\neq 0}{1\over n}\alpha_n(z^{-n}-\bar z^{-n}) \, ,
\end{array}
}
where $q$ and $p$ are the center-of-mass position and momentum for the NN string, and $q_{0}$ and $q_1$ 
denote the position of the string-endpoints of the DD string. 
Using the algebra $[\alpha_m,\alpha_n] = m\op \delta_{m+n}$, the two-point function can now be determined as 
\begin{equation}
\begin{split} 
\label{app_comp_03}
\bigl\langle \cc X_{\rm NN}(z,\bar z)\cc X_{\rm DD}(z',\bar z')\bigr\rangle =
 -{\alpha' \over 2}\Biggl(\log{z-z'\over \bar z-\bar z'}+\log{\bar z-z\over z-\bar z'}\Biggr)
 \,.
\end{split}
\end{equation}
Next, we evaluate this two-point function on the boundary $\partial\Sigma$ given by ${\rm Im}\op z=0$.
Using the same conventions as in the previous section, 
we obtain
\eq{
\bigl\langle \cc X_{\rm NN}(z,\bar z)\cc X_{\rm DD}(z',\bar z')\bigr\rangle_{\partial\Sigma} = 
-\frac{\alpha'}{2}\Biggl(\log\frac{1 + i \op y}{1 - i\op y}+\log\frac{1 - i x}{ 1 + i x}\Biggr)\,,
}
and in the limit \eqref{SWlimit} we find for the two-point function \eqref{app_comp_03} evaluated on the boundary
\eq{
\bigl\langle \cc X_{\rm NN}(\tau) \cc X_{\rm DD}(\tau')\bigr\rangle_{\partial\Sigma} = 
\frac{\pi\alpha' i }{2} \,\epsilon(\tau - \tau')\,.
}
Finally, using the two-point function, for a free theory we can 
compute the equal-time commutator as \cite{Seiberg:1999vs} 
\eq{
\label{rel_004}
\left[\cc X_{\rm NN}(\tau),\cc X_{\rm DD}(\tau)\right]= T\left(\cc X_{\rm NN}(\tau)\cc X_{\rm DD}(\tau^-)-
\cc X_{\rm NN}(\tau)\cc X_{\rm DD}(\tau^+)\right)=\pi \alpha'\op i\,.
}
Therefore, the coordinates $\cc X_{\rm NN}$ and ${\cc X}_{\rm DD}$ do not commute in the case for free fields.
We determined this commutator explicitly using the mode expansion in the next section.


\subsection{Commutator}

We now compute the commutator \eqref{rel_004} directly using the mode expansions of the 
open strings with NN and DD boundary conditions shown in \eqref{app_modes}.
Suppression the space-time index and using complex coordinates  $z=\tau+i \sigma$ on the upper half-plane, the commutator reads
\eq{
\hspace{-5pt}\left[ \cc X_{\rm NN}(\tau,\sigma),\cc X_{\rm DD} (\tau',\sigma') \right] 
&= \frac{\alpha'}{2} 
 \sum_{n\neq 0 } \frac{1}{n}\left(z^{-n}+\bar{z}^{-n}\right)\left(z'^{n}-\bar{z}'^{n}\right)
\\
&= \frac{\alpha'}{2} \sum_{n\neq 0 } \frac{1}{n}\left[ 
\left(1-\tfrac{\Delta \tau(1+i y)}{\tau+ i \sigma}\right)^n
-\left(1-\tfrac{\Delta \tau(1+i x)}{\tau+ i \sigma}\right)^n\right.\\
&\hspace{52pt}\left.
+\left(1-\tfrac{\Delta \tau(1-i x)}{\tau- i \sigma}\right)^n
-\left(1-\tfrac{\Delta \tau(1-i y)}{\tau- i \sigma}\right)^n\right] ,  \hspace{-5pt}
\label{CommPlan}
}
where we used $\Delta \tau=\tau-\tau'$ 
and employed the definition of $x$ and $y$ shown in \eqref{rel_009}. 
We are interested of the commutator for the boundary points at $\sigma=\sigma'=0$,
and apply again the limit \eqref{SWlimit}. We can therefore set $\sigma=\sigma'=\epsilon$ with $\epsilon\to0$, 
for which we find 
\begin{equation}
\begin{split} 
\left[ \cc X_{\rm NN}(\tau,\epsilon),\cc X_{\rm DD} (\tau',\epsilon) \right]\hspace{-60pt}&
\\
=\hspace{12pt}&\frac{\alpha'}{2} 
\sum_{n\neq 0 } \frac{1}{n}\left[ \left(1-\tfrac{\Delta \tau }{\tau+ i \epsilon}\right)^n
-\left(1-\tfrac{\Delta \tau }{\tau- i \epsilon }\right)^n\right.
\\
&\hspace{60pt}\left.
+\left(1-\tfrac{\Delta \tau \left(1 -i\,x\right)}{\tau- i \epsilon}\right)^n
-\left(1-\tfrac{\Delta \tau\left(1+i\,x\right)}{\tau+ i \epsilon}\right)^n
\right]
	\\
=\hspace{12pt}&  \frac{\alpha'}{2} 
\sum_{n\neq 0 } \frac{1}{n}\left[ \left(1-\tfrac{\Delta \tau }{\tau+ i \epsilon}\right)^n
-\left(1-\tfrac{\Delta \tau }{\tau- i \epsilon }\right)^n\right]
\\	
+&
\frac{\alpha'}{2}\left[\log\tfrac{\Delta \tau\left(1+i\,x\right)}{\tau+ i \epsilon}
-\log\tfrac{\Delta \tau \left(1 -i\,x\right)}{\tau- i \epsilon}\right.\\	
&\hspace{60pt} \left.
+\log\tfrac{\Delta \tau \left(1-i\,x\right)}{\Delta\tau\left(1-i\,x\right)-\tau + i \epsilon}
-\log\tfrac{\Delta \tau \left(1+i\,x\right)}{\Delta\tau\left(1+i\,x\right)-\tau- i \epsilon}
\right].
\end{split}
\end{equation}
Taking the limit \eqref{SWlimit}, the  two terms in the sum of the last expression cancel each other. The contributions of the remaining four terms depend on the sign of $\tau$ and $\tau'$: for $\tau,\tau'>0$ the terms in the parenthesis give $+2\pi i$, and for $\tau,\tau'<0$ the contribution is $-2\pi i$.  We therefore have 
\begin{equation}\label{app_rel_007}
\left[ \cc X_{\rm NN} ( \tau,0 ) ,\cc X_{\rm DD} (\tau,0 ) \right]=
\left\{
\begin{array}{@{\hspace{2pt}}l@{\hspace{20pt}}l}
 +i\pi \op\alpha'  & \text{if }\tau>0\,,\\[4pt]
 -i\pi\op \alpha'   & \text{if }\tau<0\,.
\end{array}
\right. 
\end{equation}

As a consistency check, let us also consider the situation with the end-points of two open strings on different boundaries. 
We therefore consider the limit 
\eq{
\arraycolsep1.5pt
\begin{array}{lclcl}
\sigma=\sigma' &=&\epsilon & \rightarrow& 0 \,,\\[2pt]
\tau+\tau' &=& \delta &\rightarrow&  0 \,,
\end{array}
}
which when applied to  \eqref{CommPlan} gives
\eq{
\left[ \cc X_{\rm NN}(\tau,\epsilon),\cc X_{\rm DD} (-\tau+\delta,\epsilon) \right] 
=\frac{\alpha'}{2} 
\sum_{n\neq 0 } \frac{1}{n}\Bigl[ \hspace{6pt}&\left(-\tfrac{\tau-\delta - i \epsilon}{\tau + i \epsilon}\right)^n
-\left(-\tfrac{\tau-\delta + i \epsilon}{\tau - i \epsilon}\right)^n 
\\
+&
\left(-\tfrac{\tau-\delta - i \epsilon}{\tau - i \epsilon}\right)^n
	-\left(-\tfrac{\tau-\delta + i \epsilon}{\tau + i \epsilon}\right)^n\Bigr].
}
For $\epsilon,\delta\rightarrow 0$ while keeping $\tau$ fixed and non-vanishing, 
the terms cancels each other and we obtain the expected result
\begin{equation}
\left[ \cc X_{\rm NN}(\tau,0),\cc X_{\rm DD} (-\tau,0) \right]  =0 \,.
\end{equation}  


\subsection{Commutator for open/closed-string coordinates}
\label{app_oc_com}

We now compute the commutator between the NN open-string coordinate and the dual closed-string coordinate 
using the mode expansions \eqref{app_modes} for the open string and
\begin{equation}
\begin{split} 
	\tilde X_{\text{closed}}(z,\bar z) =  q_0+ {p_0 \over 2 \pi i} \log {z\over \bar z}+\displaystyle i\sqrt{{\alpha' \over 2}}\sum_{n\neq 0}{1\over n}(\alpha_n z^{-n}-\bar \alpha_n \bar z^{-n})\,.
\end{split}
\end{equation}
for a dual closed-string coordinate.
The commutator can then be evaluated as
\begin{equation}
\begin{split} 
	\left[ \cc  X_{\rm NN}(\tau,\sigma),\tilde X_{\rm closed} (\tau',\sigma') \right] 
&= \frac{\alpha'}{2} 
 \sum_{n\neq 0 } \frac{1}{n}\left(z^{-n}+\bar{z}^{-n}\right)\left(z'^{n}-\bar{z}'^{n}\right),
\end{split}
\end{equation}
and comparing this with \eqref{CommPlan} shows that the commutators take the same form. We therefore 
find the same result as on the right-hand side of \eqref{app_rel_007}, that is
\begin{equation}
	\begin{split} 
	\left[ \cc  X_{\rm NN}(\tau,0),\tilde X_{\rm closed} (\tau',0) \right]
	=&
\left\{
\begin{array}{@{\hspace{2pt}}l@{\hspace{20pt}}l}
 +i\pi \op\alpha'  & \text{if }\tau>0\,,\\[4pt]
 -i\pi\op \alpha'   & \text{if }\tau<0\,.
\end{array}
\right. \label{CommNNdclosed}
	\end{split}
\end{equation}

  
\bibliography{references} 

\providecommand{\href}[2]{#2}\begingroup\raggedright\begin{thebibliography}{10}

\bibitem{Connes:1997cr}
A.~Connes, M.~R. Douglas, and A.~S. Schwarz, ``{Noncommutative geometry and
  matrix theory: Compactification on tori},'' {\em JHEP} {\bf 02} (1998) 003,
  \href{http://xxx.lanl.gov/abs/hep-th/9711162}{{\tt hep-th/9711162}}.

\bibitem{Douglas:1997fm}
M.~R. Douglas and C.~M. Hull, ``{D-branes and the noncommutative torus},'' {\em
  JHEP} {\bf 02} (1998) 008, \href{http://xxx.lanl.gov/abs/hep-th/9711165}{{\tt
  hep-th/9711165}}.

\bibitem{Chu:1998qz}
C.-S. Chu and P.-M. Ho, ``{Noncommutative open string and D-brane},'' {\em
  Nucl. Phys.} {\bf B550} (1999) 151--168,
  \href{http://xxx.lanl.gov/abs/hep-th/9812219}{{\tt hep-th/9812219}}.

\bibitem{Schomerus:1999ug}
V.~Schomerus, ``{D-branes and deformation quantization},'' {\em JHEP} {\bf 06}
  (1999) 030, \href{http://xxx.lanl.gov/abs/hep-th/9903205}{{\tt
  hep-th/9903205}}.

\bibitem{Seiberg:1999vs}
N.~Seiberg and E.~Witten, ``{String theory and noncommutative geometry},'' {\em
  JHEP} {\bf 09} (1999) 032, \href{http://xxx.lanl.gov/abs/hep-th/9908142}{{\tt
  hep-th/9908142}}.

\bibitem{Szabo:2001kg}
R.~J. Szabo, ``{Quantum field theory on noncommutative spaces},'' {\em Phys.
  Rept.} {\bf 378} (2003) 207--299,
  \href{http://xxx.lanl.gov/abs/hep-th/0109162}{{\tt hep-th/0109162}}.

\bibitem{Cornalba:2001sm}
L.~Cornalba and R.~Schiappa, ``{Nonassociative star product deformations for
  D-brane world volumes in curved backgrounds},'' {\em Commun. Math. Phys.}
  {\bf 225} (2002) 33--66, \href{http://xxx.lanl.gov/abs/hep-th/0101219}{{\tt
  hep-th/0101219}}.

\bibitem{Herbst:2001ai}
M.~Herbst, A.~Kling, and M.~Kreuzer, ``{Star products from open strings in
  curved backgrounds},'' {\em JHEP} {\bf 09} (2001) 014,
  \href{http://xxx.lanl.gov/abs/hep-th/0106159}{{\tt hep-th/0106159}}.

\bibitem{Plauschinn:2018wbo}
E.~Plauschinn, ``{Non-geometric backgrounds in string theory},'' {\em Phys.
  Rept.} {\bf 798} (2019) 1--122,
  \href{http://xxx.lanl.gov/abs/1811.11203}{{\tt 1811.11203}}.

\bibitem{Blumenhagen:2010hj}
R.~Blumenhagen and E.~Plauschinn, ``{Nonassociative Gravity in String
  Theory?},'' {\em J. Phys.} {\bf A44} (2011) 015401,
  \href{http://xxx.lanl.gov/abs/1010.1263}{{\tt 1010.1263}}.

\bibitem{Lust:2010iy}
D.~L{\"u}st, ``{T-duality and closed string non-commutative (doubled)
  geometry},'' {\em JHEP} {\bf 1012} (2010) 084,
  \href{http://xxx.lanl.gov/abs/1010.1361}{{\tt 1010.1361}}.

\bibitem{Blumenhagen:2011ph}
R.~Blumenhagen, A.~Deser, D.~L{\"u}st, E.~Plauschinn, and F.~Rennecke,
  ``{Non-geometric Fluxes, Asymmetric Strings and Nonassociative Geometry},''
  {\em J. Phys.} {\bf A44} (2011) 385401,
  \href{http://xxx.lanl.gov/abs/1106.0316}{{\tt 1106.0316}}.

\bibitem{Blumenhagen:2011yv}
R.~Blumenhagen, ``{Nonassociativity in String Theory},'' in {\em Strings, gauge
  fields, and the geometry behind: The legacy of Maximilian Kreuzer}
  (A.~Rebhan, L.~Katzarkov, J.~Knapp, R.~Rashkov, and E.~Scheidegger, eds.),
  pp.~213--224.
\newblock 2011.
\newblock \href{http://xxx.lanl.gov/abs/1112.4611}{{\tt 1112.4611}}.

\bibitem{Lust:2012fp}
D.~L{\"u}st, ``{Twisted Poisson Structures and Non-commutative/non-associative
  Closed String Geometry},'' {\em PoS} {\bf CORFU2011} (2011) 086,
  \href{http://xxx.lanl.gov/abs/1205.0100}{{\tt 1205.0100}}.

\bibitem{Bakas:2013jwa}
I.~Bakas and D.~L{\"u}st, ``{3-Cocycles, Non-Associative Star-Products and the
  Magnetic Paradigm of R-Flux String Vacua},'' {\em JHEP} {\bf 01} (2014) 171,
  \href{http://xxx.lanl.gov/abs/1309.3172}{{\tt 1309.3172}}.

\bibitem{Condeescu:2012sp}
C.~Condeescu, I.~Florakis, and D.~L{\"u}st, ``{Asymmetric Orbifolds,
  Non-Geometric Fluxes and Non-Commutativity in Closed String Theory},'' {\em
  JHEP} {\bf 04} (2012) 121, \href{http://xxx.lanl.gov/abs/1202.6366}{{\tt
  1202.6366}}.

\bibitem{Andriot:2012vb}
D.~Andriot, M.~Larfors, D.~L{\"u}st, and P.~Patalong, ``{(Non-) commutative
  closed string on T-dual toroidal backgrounds},'' {\em JHEP} {\bf 1306} (2013)
  021, \href{http://xxx.lanl.gov/abs/1211.6437}{{\tt 1211.6437}}.

\bibitem{Blair:2014kla}
C.~D.~A. Blair, ``{Non-commutativity and non-associativity of the doubled
  string in non-geometric backgrounds},'' {\em JHEP} {\bf 1506} (2015) 091,
  \href{http://xxx.lanl.gov/abs/1405.2283}{{\tt 1405.2283}}.

\bibitem{Bakas:2015gia}
I.~Bakas and D.~L{\"u}st, ``{T-duality, Quotients and Currents for
  Non-Geometric Closed Strings},'' {\em Fortsch. Phys.} {\bf 63} (2015)
  543--570, \href{http://xxx.lanl.gov/abs/1505.04004}{{\tt 1505.04004}}.

\bibitem{Bouwknegt:2000qt}
P.~Bouwknegt and V.~Mathai, ``{D-branes, B fields and twisted K theory},'' {\em
  JHEP} {\bf 03} (2000) 007, \href{http://xxx.lanl.gov/abs/hep-th/0002023}{{\tt
  hep-th/0002023}}.

\bibitem{Bouwknegt:2004ap}
P.~Bouwknegt, K.~Hannabuss, and V.~Mathai, ``{Nonassociative tori and
  applications to T-duality},'' {\em Commun. Math. Phys.} {\bf 264} (2006)
  41--69, \href{http://xxx.lanl.gov/abs/hep-th/0412092}{{\tt hep-th/0412092}}.

\bibitem{Mathai:2004qq}
V.~Mathai and J.~M. Rosenberg, ``{T duality for torus bundles with H fluxes via
  noncommutative topology},'' {\em Commun. Math. Phys.} {\bf 253} (2004)
  705--721, \href{http://xxx.lanl.gov/abs/hep-th/0401168}{{\tt
  hep-th/0401168}}.

\bibitem{Mathai:2004qc}
V.~Mathai and J.~M. Rosenberg, ``{On Mysteriously missing T-duals, H-flux and
  the T-duality group},'' in {\em {Differential geometry and physics.
  Proceedings, 23rd International Conference, Tianjin, China, August 20-26,
  2005}}, pp.~350--358, 2004.
\newblock \href{http://xxx.lanl.gov/abs/hep-th/0409073}{{\tt hep-th/0409073}}.

\bibitem{Mylonas:2012pg}
D.~Mylonas, P.~Schupp, and R.~J. Szabo, ``{Membrane Sigma-Models and
  Quantization of Non-Geometric Flux Backgrounds},'' {\em JHEP} {\bf 1209}
  (2012) 012, \href{http://xxx.lanl.gov/abs/1207.0926}{{\tt 1207.0926}}.

\bibitem{Freidel:2017wst}
L.~Freidel, R.~G. Leigh, and D.~Minic, ``{Intrinsic non-commutativity of closed
  string theory},'' {\em JHEP} {\bf 09} (2017) 060,
  \href{http://xxx.lanl.gov/abs/1706.03305}{{\tt 1706.03305}}.

\bibitem{Freidel:2017nhg}
L.~Freidel, R.~G. Leigh, and D.~Minic, ``{Noncommutativity of closed string
  zero modes},'' {\em Phys. Rev.} {\bf D96} (2017), no.~6 066003,
  \href{http://xxx.lanl.gov/abs/1707.00312}{{\tt 1707.00312}}.

\bibitem{Blumenhagen:2000wh}
R.~Blumenhagen, L.~G{\"o}rlich, B.~K{\"o}rs, and D.~L{\"u}st, ``{Noncommutative
  compactifications of type I strings on tori with magnetic background flux},''
  {\em JHEP} {\bf 10} (2000) 006,
  \href{http://xxx.lanl.gov/abs/hep-th/0007024}{{\tt hep-th/0007024}}.

\bibitem{Blumenhagen:2000fp}
R.~Blumenhagen, L.~G{\"o}rlich, B.~K{\"o}rs, and D.~L{\"u}st, ``{Asymmetric
  orbifolds, noncommutative geometry and type I string vacua},'' {\em Nucl.
  Phys.} {\bf B582} (2000) 44--64,
  \href{http://xxx.lanl.gov/abs/hep-th/0003024}{{\tt hep-th/0003024}}.

\bibitem{Blumenhagen:2018kwq}
R.~Blumenhagen, I.~Brunner, V.~Kupriyanov, and D.~L{\"u}st, ``{Bootstrapping
  non-commutative gauge theories from L$_\infty$ algebras},'' {\em JHEP} {\bf
  05} (2018) 097, \href{http://xxx.lanl.gov/abs/1803.00732}{{\tt 1803.00732}}.

\bibitem{Grange:2006es}
P.~Grange and S.~Schafer-Nameki, ``{T-duality with H-flux: Non-commutativity,
  T-folds and G x G structure},'' {\em Nucl. Phys.} {\bf B770} (2007) 123--144,
  \href{http://xxx.lanl.gov/abs/hep-th/0609084}{{\tt hep-th/0609084}}.

\bibitem{Brodzki:2007hg}
J.~Brodzki, V.~Mathai, J.~M. Rosenberg, and R.~J. Szabo, ``{Noncommutative
  correspondences, duality and D-branes in bivariant K-theory},'' {\em Adv.
  Theor. Math. Phys.} {\bf 13} (2009), no.~2 497--552,
  \href{http://xxx.lanl.gov/abs/0708.2648}{{\tt 0708.2648}}.

\bibitem{Kapustin:1999di}
A.~Kapustin, ``{D-branes in a topologically nontrivial B field},'' {\em Adv.
  Theor. Math. Phys.} {\bf 4} (2000) 127--154,
  \href{http://xxx.lanl.gov/abs/hep-th/9909089}{{\tt hep-th/9909089}}.

\bibitem{Mylonas:2013jha}
D.~Mylonas, P.~Schupp, and R.~J. Szabo, ``{Non-Geometric Fluxes, Quasi-Hopf
  Twist Deformations and Nonassociative Quantum Mechanics},'' {\em J. Math.
  Phys.} {\bf 55} (2014) 122301, \href{http://xxx.lanl.gov/abs/1312.1621}{{\tt
  1312.1621}}.

\bibitem{Freed:1999vc}
D.~S. Freed and E.~Witten, ``{Anomalies in string theory with D-branes},'' {\em
  Asian J. Math.} {\bf 3} (1999) 819,
  \href{http://xxx.lanl.gov/abs/hep-th/9907189}{{\tt hep-th/9907189}}.

\bibitem{Hull:2019iuy}
C.~Hull and R.~J. Szabo, ``{Noncommutative gauge theories on D-branes in
  non-geometric backgrounds},'' {\em JHEP} {\bf 09} (2019) 051,
  \href{http://xxx.lanl.gov/abs/1903.04947}{{\tt 1903.04947}}.

\bibitem{Fradkin:1985qd}
E.~S. Fradkin and A.~A. Tseytlin, ``{Nonlinear Electrodynamics from Quantized
  Strings},'' {\em Phys. Lett.} {\bf 163B} (1985) 123--130.

\bibitem{Abouelsaood:1986gd}
A.~Abouelsaood, C.~G. Callan, Jr., C.~R. Nappi, and S.~A. Yost, ``{Open Strings
  in Background Gauge Fields},'' {\em Nucl. Phys.} {\bf B280} (1987) 599--624.

\bibitem{Callan:1986bc}
C.~G. Callan, Jr., C.~Lovelace, C.~R. Nappi, and S.~A. Yost, ``{String Loop
  Corrections to beta Functions},'' {\em Nucl. Phys.} {\bf B288} (1987)
  525--550.

\bibitem{Dasgupta:1999ss}
K.~Dasgupta, G.~Rajesh, and S.~Sethi, ``{M theory, orientifolds and G -
  flux},'' {\em JHEP} {\bf 08} (1999) 023,
  \href{http://xxx.lanl.gov/abs/hep-th/9908088}{{\tt hep-th/9908088}}.

\bibitem{Kachru:2002sk}
S.~Kachru, M.~B. Schulz, P.~K. Tripathy, and S.~P. Trivedi, ``{New
  supersymmetric string compactifications},'' {\em JHEP} {\bf 03} (2003) 061,
  \href{http://xxx.lanl.gov/abs/hep-th/0211182}{{\tt hep-th/0211182}}.

\bibitem{Hull:2004in}
C.~Hull, ``{A Geometry for non-geometric string backgrounds},'' {\em JHEP} {\bf
  0510} (2005) 065, \href{http://xxx.lanl.gov/abs/hep-th/0406102}{{\tt
  hep-th/0406102}}.

\bibitem{Shelton:2005cf}
J.~Shelton, W.~Taylor, and B.~Wecht, ``{Nongeometric flux compactifications},''
  {\em JHEP} {\bf 0510} (2005) 085,
  \href{http://xxx.lanl.gov/abs/hep-th/0508133}{{\tt hep-th/0508133}}.

\bibitem{Lust:2017jox}
D.~Lüst, E.~Plauschinn, and V.~Vall~Camell, ``{Unwinding strings in
  semi-flatland},'' {\em JHEP} {\bf 07} (2017) 027,
  \href{http://xxx.lanl.gov/abs/1706.00835}{{\tt 1706.00835}}.

\bibitem{Cordonier-Tello:2018zdw}
F.~Cordonier-Tello, D.~L{\"u}st, and E.~Plauschinn, ``{Open-string T-duality
  and applications to non-geometric backgrounds},'' {\em JHEP} {\bf 08} (2018)
  198, \href{http://xxx.lanl.gov/abs/1806.01308}{{\tt 1806.01308}}.

\bibitem{Dabholkar:2005ve}
A.~Dabholkar and C.~Hull, ``{Generalised T-duality and non-geometric
  backgrounds},'' {\em JHEP} {\bf 0605} (2006) 009,
  \href{http://xxx.lanl.gov/abs/hep-th/0512005}{{\tt hep-th/0512005}}.

\bibitem{Hull:2006va}
C.~M. Hull, ``{Doubled Geometry and T-Folds},'' {\em JHEP} {\bf 0707} (2007)
  080, \href{http://xxx.lanl.gov/abs/hep-th/0605149}{{\tt hep-th/0605149}}.

\bibitem{Hull:2009sg}
C.~Hull and R.~Reid-Edwards, ``{Non-geometric backgrounds, doubled geometry and
  generalised T-duality},'' {\em JHEP} {\bf 0909} (2009) 014,
  \href{http://xxx.lanl.gov/abs/0902.4032}{{\tt 0902.4032}}.

\bibitem{Hitchin:2004ut}
N.~Hitchin, ``{Generalized Calabi-Yau manifolds},'' {\em Quart.J.Math.Oxford
  Ser.} {\bf 54} (2003) 281--308,
  \href{http://xxx.lanl.gov/abs/math/0209099}{{\tt math/0209099}}.

\bibitem{Hitchin:2005in}
N.~Hitchin, ``{Brackets, forms and invariant functionals},''
  \href{http://xxx.lanl.gov/abs/math/0508618}{{\tt math/0508618}}.

\bibitem{Hull:2009mi}
C.~Hull and B.~Zwiebach, ``{Double Field Theory},'' {\em JHEP} {\bf 0909}
  (2009) 099, \href{http://xxx.lanl.gov/abs/0904.4664}{{\tt 0904.4664}}.

\bibitem{Hohm:2010pp}
O.~Hohm, C.~Hull, and B.~Zwiebach, ``{Generalized metric formulation of double
  field theory},'' {\em JHEP} {\bf 1008} (2010) 008,
  \href{http://xxx.lanl.gov/abs/1006.4823}{{\tt 1006.4823}}.

\bibitem{Hohm:2012mf}
O.~Hohm and B.~Zwiebach, ``{Towards an invariant geometry of double field
  theory},'' {\em J.Math.Phys.} {\bf 54} (2013) 032303,
  \href{http://xxx.lanl.gov/abs/1212.1736}{{\tt 1212.1736}}.

\bibitem{Hohm:2013bwa}
O.~Hohm, D.~L{\"u}st, and B.~Zwiebach, ``{The Spacetime of Double Field Theory:
  Review, Remarks, and Outlook},'' {\em Fortsch.Phys.} {\bf 61} (2013)
  926--966, \href{http://xxx.lanl.gov/abs/1309.2977}{{\tt 1309.2977}}.

\bibitem{Andriot:2011uh}
D.~Andriot, M.~Larfors, D.~L{\"u}st, and P.~Patalong, ``{A ten-dimensional
  action for non-geometric fluxes},'' {\em JHEP} {\bf 1109} (2011) 134,
  \href{http://xxx.lanl.gov/abs/1106.4015}{{\tt 1106.4015}}.

\bibitem{Andriot:2012wx}
D.~Andriot, O.~Hohm, M.~Larfors, D.~L{\"u}st, and P.~Patalong, ``{A geometric
  action for non-geometric fluxes},'' {\em Phys.Rev.Lett.} {\bf 108} (2012)
  261602, \href{http://xxx.lanl.gov/abs/1202.3060}{{\tt 1202.3060}}.

\bibitem{Andriot:2012an}
D.~Andriot, O.~Hohm, M.~Larfors, D.~L{\"u}st, and P.~Patalong, ``{Non-Geometric
  Fluxes in Supergravity and Double Field Theory},'' {\em Fortsch.Phys.} {\bf
  60} (2012) 1150--1186, \href{http://xxx.lanl.gov/abs/1204.1979}{{\tt
  1204.1979}}.

\bibitem{Blumenhagen:2012nk}
R.~Blumenhagen, A.~Deser, E.~Plauschinn, and F.~Rennecke, ``{A bi-invariant
  Einstein-Hilbert action for the non-geometric string},'' {\em Phys.Lett.}
  {\bf B720} (2013) 215--218, \href{http://xxx.lanl.gov/abs/1210.1591}{{\tt
  1210.1591}}.

\bibitem{Blumenhagen:2012nt}
R.~Blumenhagen, A.~Deser, E.~Plauschinn, and F.~Rennecke, ``{Non-geometric
  strings, symplectic gravity and differential geometry of Lie algebroids},''
  {\em JHEP} {\bf 1302} (2013) 122,
  \href{http://xxx.lanl.gov/abs/1211.0030}{{\tt 1211.0030}}.

\bibitem{Blumenhagen:2013aia}
R.~Blumenhagen, A.~Deser, E.~Plauschinn, F.~Rennecke, and C.~Schmid, ``{The
  Intriguing Structure of Non-geometric Frames in String Theory},'' {\em
  Fortsch. Phys.} {\bf 61} (2013) 893--925,
  \href{http://xxx.lanl.gov/abs/1304.2784}{{\tt 1304.2784}}.

\bibitem{Hohm:2017cey}
O.~Hohm, V.~Kupriyanov, D.~L{\"u}st, and M.~Traube, ``{Constructions of
  $L_{\infty}$ algebras and their field theory realizations},'' {\em Adv. Math.
  Phys.} {\bf 2018} (2018) 9282905,
  \href{http://xxx.lanl.gov/abs/1709.10004}{{\tt 1709.10004}}.

\bibitem{Blumenhagen:2018shf}
R.~Blumenhagen, M.~Brinkmann, V.~Kupriyanov, and M.~Traube, ``{On the
  Uniqueness of L$_\infty$ bootstrap: Quasi-isomorphisms are Seiberg-Witten
  Maps},'' {\em J. Math. Phys.} {\bf 59} (2018), no.~12 123505,
  \href{http://xxx.lanl.gov/abs/1806.10314}{{\tt 1806.10314}}.

\end{thebibliography}\endgroup
\bibliographystyle{utphys}


\end{document}